\documentclass[a4paper]{spie}%
\usepackage{graphicx}
\usepackage{amsmath}
\usepackage{amsfonts}
\usepackage{amssymb}
\usepackage{subfigure}%
\begin{document}

\title{Improving the lower bound to the secret-key capacity\\of the thermal amplifier channel}
\author{Gan Wang\supit{a,b}, Carlo Ottaviani\supit{b}, Hong Guo\supit{a}, and Stefano
Pirandola\supit{b} \skiplinehalf
\supit{a}State Key Laboratory of Advanced Optical Communication Systems and
Networks, School of Electronics Engineering and Computer Science, and Center
for Quantum Information Technology, Peking University, Beijing 100871, China;\\\supit{b}Computer Science, University of York, York YO10 5GH, United Kingdom.}
\maketitle

\begin{abstract}
We consider the noisy thermal amplifier channel, where signal modes are
amplified together with environmental thermal modes. We focus on the
secret-key capacity of this channel, which is the maximum amount of secret
bits that two remote parties can generate by means of the most general
adaptive protocol, assisted by unlimited and two-way classical communication.
For this channel only upper and lower bounds are known, and in this work we
improve the lower bound. We consider a protocol based on squeezed states and
homodyne detections, in both direct and reverse reconciliation. In particular,
we assume that trusted thermal noise is mixed on beam splitters controlled by
the parties in a way to assist their homodyne detections. The new improved
lower bounds to the secret-key capacity are obtained by optimizing the key
rates over the variance of the trusted noise injected, and the transmissivity
of the parties' beam splitters. Our results confirm that there is a separation
between the coherent information of the thermal amplifier channel and its
secret key capacity.

\end{abstract}


\section{Introduction}

In the past decades, quantum information science\cite{Watrous,Hayashi} has
successfully achieved a huge amount of goals. In particular quantum key
distribution (QKD) has emerged as the most mature quantum technology. The aim
of QKD is to distribute secret keys between two parties, a sender (Alice) and
a receiver (Bob), who perform a communication scheme in two stages. The first
stage is quantum communication over a quantum channel controlled by an
eavesdropper (Eve), ending with Alice and Bob sharing a \emph{raw key}. During
the second stage of classical communication, the parties run a classical
protocol of error correction, sifting and privacy amplification. In this way
they extract a shorter key over which Eve only has a negligible amount of
knowledge. The fundamental mechanism ensuring security is the no-cloning
theorem \cite{nocloning}, which forbids a perfect copy of the non-orthogonal
signal states sent by Alice.

Two main designs of QKD exist. One is based on qubits \cite{Scarani2009}, the
other is based on continuous-variable (CV) quantum systems \cite{SAM-RMP,RMP},
which are described by an infinite-dimensional Hilbert space. In particular,
Gaussian CV QKD \cite{RMP} received a lot of attention for the relative
simplicity of its theoretical analysis, and the simplicity of its experimental
realization based on cheap, \emph{off-the-shelf}, linear optical elements and
highly efficient homodyne detectors, even at non-standard
frequencies~\cite{furusawa2018}. During the past years, the research in
Gaussian QKD has led to the design and experimental implementation of a number
of protocols, including
one-way\cite{GrosshansNAT,Chris1,Chris2,Lutkenaus,PC-PRL2009,Thermal1,Thermal2,Thermal3,Thermal4,Thermal5,Thermal6,Usenko,prova1,prova2,prova3}%
, and
two-way\cite{pirs2way,Ott2way1,Ott2way2,Thermal2way,Jeff1,Jeff2,Jeff3,Ghorai}
schemes, as well as the study of
measurement-device-independent\cite{CVMDIQKD,lupo2018parameter,PRA-S,li2014continuous,zhang2014continuous,papanastasiou2017finite,lupo2018continuous}
schemes.

An important goal in this research area is to determine the optimal secret key
rate, or secret-key capacity, over the various models of quantum communication
channels. This computation is generally complicated due to the fact that
feedback has to be taken into account. More precisely, one has to optimize the
key-rate over adaptive LOCCs, i.e., local operations (LOs) assisted by
unlimited two-way classical communication (CC). The combined use of the
relative entropy of entanglement (REE) and teleportation stretching allowed
PLOB\cite{PLOB} to upperbound the secret-key capacities of Pauli channels,
erasure channels, amplitude damping channels, and bosonic Gaussian channels
(see also the follow-up works~\cite{follow1,follow2,follow3,follow4,follow5}).
Among the Gaussian channels, the thermal loss channel and the thermal
amplifier are the most interesting and important. In a previous
work\cite{ottaviani2016secret}, we showed how the lower bound to the secret
key capacity of the thermal loss channel can be improved by exploiting the
benefits of injecting trusted thermal noise\cite{PC-PRL2009,Thermal6}. This
type of\ analysis has not been yet performed for the thermal amplifier channel.

In this work, we improve the lower bound to the secret-key capacity of the
thermal amplifier channel by computing the achievable rate of a QKD\ protocol
based on squeezed states and homodyne detections. We assume that the parties
possess quantum memory so that they do not need to reconcilate their bases,
i.e., the choices of the $q$ or the $p$ quadrature. We also assume that
trusted thermal noise is locally used by Alice or Bob, depending on whether
the protocol is implemented in direct reconcilation (DR) or reverse
reconciliation (RR). Under these conditions, the lower bound based on the
coherent information~\cite{cohe1,cohe2} is always beaten by the rate in DR
(and also outperformed by the RR rate in a small region for low gains and high
thermal noise).

\section{Upper and lower bounds to the secret-key capacity of the thermal
amplifier}

Consider two parties, Alice and Bob, performing an adaptive protocol over a
quantum channel $\mathcal{E}$. After $n$ uses, they share the output state
$\rho_{n}:=\rho\left(  \mathcal{E}^{\otimes n}\right)  $ which depends on the
sequence of adaptive LOCCs performed, i.e., $\mathcal{L}=\{\Lambda_{0}%
,\Lambda_{1},...,\Lambda_{n}\}$. Let $\phi_{n}$ be a private target state
\cite{private} with information content equal to $nR_{n}$ secret bits. The
output state $\rho_{n}$ and $\phi_{n}$ fulfill the $\epsilon$-security
relation $||\rho_{n}-\phi_{n}||\leq\epsilon$. Now, the generic two-way
capacity of the channel can be obtained by optimizing over all the possible
LOCC-sequences $\mathcal{L}$, and by taking the limit of infinite channel
uses, i.e., $n\rightarrow\infty$. In formulas we can define the secret key
capacity as follows
\begin{equation}
\mathcal{K}(\mathcal{E}):=\sup\limits_{\mathcal{L}}\lim\limits_{n\rightarrow
\infty}R_{n}. \label{capacity}%
\end{equation}
This quantity gives the maximum achievable number of secret bits that can be
transmitted per channel use.

Let us introduce the quadrature vector $\hat{\mathbf{x}}:=(q,p)^{T}$. Then, a
thermal amplifier channel $\mathcal{E}_{g,\bar{n}}$ corresponds to the
transformation
\begin{equation}
\hat{\mathbf{x}}\rightarrow\sqrt{g}\hat{\mathbf{x}}+\sqrt{g-1}\hat{\mathbf{x}%
}_{E}%
\end{equation}
where $g>1$ is the gain, and $\hat{\mathbf{x}}_{E}$ are the quadratures of a
thermal environment mode $E$ with $\bar{n}$ mean number of photons. Let us set
$\omega=2\bar{n}+1$ and
\begin{equation}
h(x):=\frac{x+1}{2}\log_{2}\frac{x+1}{2}-\frac{x-1}{2}\log_{2}\frac{x-1}{2}.
\end{equation}
Then, we may write the secret-key capacity of the thermal amplifier channel
$\mathcal{K}(\mathcal{E}_{g,\bar{n}})$ as%
\begin{equation}
\Omega(g,\bar{n})\leq\mathcal{K}(\mathcal{E}_{g,\bar{n}})\leq\Phi(g,\bar
{n}),\label{LB-UB}%
\end{equation}
where the lower bound\cite{holevo2001evaluating} is given by%
\begin{equation}
\Omega(g,\bar{n})=\log_{2}\left(  \frac{g}{g-1}\right)  -h(\omega
)\label{LBonly}%
\end{equation}
and corresponds to the coherent information of the channel, which is defined
as the coherent information of its (asymptotic) Choi
matrix\cite{PLOB,pirandola2009direct}. In Eq.~(\ref{LB-UB}), the upper bound
is computed from the REE~\cite{REE1,REE2,REE3} of the (asymptotic) Choi matrix
and is equal to\cite{PLOB}
\begin{equation}
\Phi(g,\bar{n})=\left\{
\begin{array}
[c]{l}%
\log_{2}\left(  \frac{g^{\bar{n}+1}}{g-1}\right)  -h(\omega),~\text{for}%
~\bar{n}<(g-1)^{-1}\\
\\
0,~\text{otherwise.}%
\end{array}
\right.  \label{UBonly}%
\end{equation}

\section{Improving the lower bound}

We now present a QKD\ protocol whose key rate in DR and RR improves the lower
bound in Eq.~(\ref{LB-UB}). Even though the improvement found is small, it is
meaningful because it shows that the coherent information of the thermal
amplifier channel is cannot be its secret key capacity. First we derive the
new achievable rates in subsections~\ref{SECDR} and~\ref{SECRR}. Then we
numercally compare the results in subsection~\ref{SecCOMP}.

\subsection{Achievable rate in direct reconciliation\label{SECDR}}

We show the following result.

\begin{theorem}
Consider a thermal amplifier channel, with gain $g$ and thermal noise $\omega
$. Its secret key rate is lower-bounded by the achievable DR\ rate
\begin{equation}
R^{\blacktriangleright}(g,\omega)=\underset{\eta_{A},\gamma}{\max
}~R^{\blacktriangleright}\left(  g,\omega,\eta_{A},\gamma\right)  ,
\label{Bound-DR}%
\end{equation}
where
\begin{equation}
R^{\blacktriangleright}(g,\omega,\eta_{A},\gamma):=\frac{1}{2}\log_{2}%
\frac{g[g\eta_{A}\omega+\gamma(g-1)(1-\eta_{A})]}{(g-1)[g\gamma(1-\eta
_{A})+\eta_{A}\omega(g-1)]}+h\left(  \sqrt{\frac{\omega\lbrack g\eta
_{A}+\gamma\omega(g-1)(1-\eta_{A})]}{g\eta_{A}\omega+\gamma(g-1)(1-\eta_{A})}%
}\right)  -h(\omega), \label{Rate-DR}%
\end{equation}
and the maximization is over transmissivity $\eta_{A}$ of a beam splitter at
Alice's side, and the thermal variance $\gamma\geq1$.
\end{theorem}

\begin{figure}[th]
\centering\includegraphics[width=13cm]{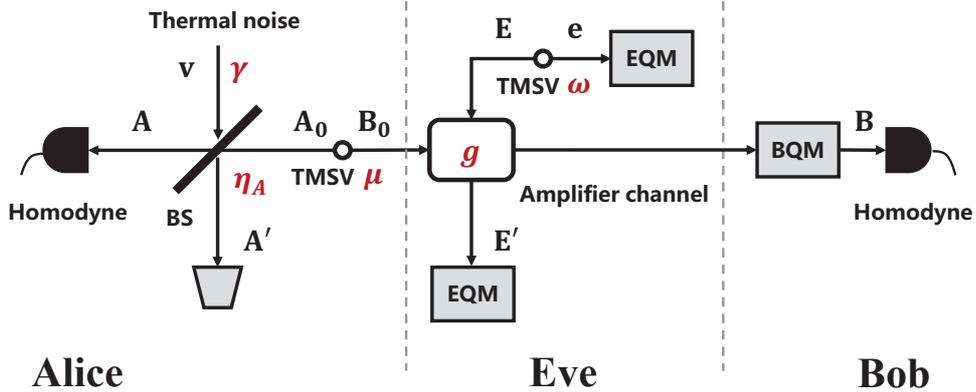} \caption{Protocol
with trusted thermal noise in DR. Alice has a TMSV state, whose
mode $B_{0}$ is sent to Bob through a thermal amplifier channel
with gain $g$. Mode $A_{0}$ is processed by a beam splitter (BS)
with transmissivity $\eta_{A}$ together with a thermal mode $v$
with variance $\gamma$, and then measured by a homodyne detector
in $q$ or $p$. The attack is performed by Eve, who exploits modes
$e$ and $E$ in a TMSV state (with variance $\omega$) and stores
the output in Eve's quantum memory (EQM). The signals from Alice
are stored by Bob in his quantum memory (BQM) and homodyned after
Alice has sent to Bob the correct sequence of
homodyne detections to perform.}%
\label{Structure-DR}%
\end{figure}

\textbf{Proof.} Consider the Gaussian CV-QKD protocol described in
Fig.~\ref{Structure-DR}. We study its security in the entanglement-based (EB)
representation. Thus, we assume that Alice has a two-mode squeezed vacuum
(TMSV) state $\Phi^{\mu}$ of modes $A_{0}$ and $B_{0}$. The covariance matrix
(CM) describing this zero mean Gaussian state is the following \cite{RMP}%
\begin{equation}
\mathbf{V}_{A_{0}B_{0}}=\mathbf{V}_{\mathrm{TMSV}}(\mu):=\left(
\begin{matrix}
\mu\mathbf{I} & \sqrt{\mu^{2}-1}\mathbf{Z}\\
\sqrt{\mu^{2}-1}\mathbf{Z} & \mu\mathbf{I}%
\end{matrix}
\right)  , \label{VA0B0}%
\end{equation}
where $\mathbf{I=}$\textrm{diag}$(1,1)$ and $\mathbf{Z=}$\textrm{diag}%
$(1,-1)$, and $\mu$ is the variance of the TMSV state. Alice's local mode
$A_{0}$ is processed by a beam splitter with transmissivity $\eta_{A}$,
together with mode $v$ in a thermal state of variance $\gamma$, and CM
$\mathbf{V}_{v}=\gamma\mathbf{I}$. One of the outputs, $A^{\prime}$, is
discarded, while the other, $A$, is homodyned randomly switching between
quadrature $q$ and $p$. This operation prepares thermal states in the
travelling mode $B_{0}$.

Mode $B_{0}$ is sent through the thermal amplifier channel with gain $g>1$ and
thermal noise $\omega=2\bar{n}+1$. The thermal input mode $E$ is part of Eve's
TMSV\ state with CM $\mathbf{V}_{eE}=\mathbf{V}_{\mathrm{TMSV}}(\omega)$ as in
Fig.~\ref{Structure-DR}. Eve's output modes $e$ and $E^{\prime}$ are stored in
a quantum memory, which is coherently measured at the end of the protocol
(collective attack). The channel output $B$, travelling to Bob, is stored in
Bob's quantum memory (BQM) for later measurements. After many uses of the
channel ($n\gg1$), Alice communicates which quadrature she has measured in
each round, thus Bob can perform exactly the same sequence of homodyne
detections on the stored modes, and then infer the outcomes of Alice's
preparation stage.

The initial global state $\rho_{0}$ of Alice, Bob and Eve is given by the
tensor product $\rho_{0}=\rho_{A_{0}B_{0}}\otimes\rho_{eE}\otimes\rho_{v}$,
having CM $\mathbf{V}_{0}^{\blacktriangleright}=\mathbf{V}_{A_{0}B_{0}%
}\mathbf{\oplus V}_{eE}\mathbf{\oplus V}_{v}=\mathbf{V}_{A_{0}B_{0}%
eEv}^{\blacktriangleright}$. For convenience, we rearrange the state as
$\mathbf{V}_{0}^{\blacktriangleright}=\mathbf{V}_{A_{0}vB_{0}Ee}$, and process
it by a sequence of symplectic transformation describing the evolution
throughout the beam splitter ($\eta_{A}$) and the amplifier ($g$). We first
process mode $A_{0}$ and $v$, by applying the symplectic transformation
$\mathbf{\tilde{V}}^{\blacktriangleright}=\mathbf{S}_{\eta_{A}}\mathbf{V}%
_{0}\mathbf{S}_{\eta_{A}}^{T}$, where $\mathbf{S}_{\eta_{A}}:=\mathbf{T}%
_{BS}(\eta_{A})\oplus\mathbf{I\oplus I\oplus I}$, with
\begin{equation}
\mathbf{T}_{BS}(\eta_{A}):=\left(
\begin{matrix}
\sqrt{\eta_{A}}\mathbf{I} & \sqrt{1-\eta_{A}}\mathbf{I}\\
-\sqrt{1-\eta_{A}}\mathbf{I} & \sqrt{\eta_{A}}\mathbf{I}%
\end{matrix}
\right)  . \label{T-BS}%
\end{equation}
Then, we process the CM $\mathbf{\tilde{V}}^{\blacktriangleright}$ to consider
the evolution of the state through the thermal amplifier, by applying the
symplectic transformation $\mathbf{S}_{g}\mathbf{\tilde{V}}%
^{\blacktriangleright}\mathbf{S}_{g}^{T}$, where $\mathbf{S}_{g}%
:=\mathbf{I\oplus I\oplus T}_{AMP}(g)\oplus\mathbf{I}$, and
\begin{equation}
\mathbf{T}_{AMP}(g):=\left(
\begin{matrix}
\sqrt{g}\mathbf{I} & \sqrt{g-1}\mathbf{Z}\\
\sqrt{g-1}\mathbf{Z} & \sqrt{g}\mathbf{I}%
\end{matrix}
\right)  . \label{T-AMP}%
\end{equation}
Thus, we can compute the CM $\mathbf{V}^{\blacktriangleright}$, corresponding
to the quantum state $\rho_{AA^{\prime}BE^{\prime}e}$. Then we can trace out
mode $A^{\prime}$ to obtain the output state $\rho_{ABE^{\prime}e}%
=\mathrm{Tr}_{A^{\prime}}(\rho_{AA^{\prime}BE^{\prime}e})$ with CM
$\mathbf{V}_{ABE^{\prime}e}^{\blacktriangleright}$. From this CM we may
compute Alice's and Bob's mutual information $I_{AB}$ as well as Eve's Holevo
function $\chi_{AE}$, bounding Eve's knowledge on Alice's encoding variables.

Under ideal conditions of perfect reconciliation efficiency, the key rate in
DR is given by $R^{\blacktriangleright}:=I_{AB}-\chi_{AE}$. We can derive the
analytical expression of the asymptotic key rate, when the Gaussian modulation
is large $\mu\rightarrow\infty$. To compute $I_{AB}$, let us first consider
the CM describing modes $A$ and $B$. This is given by the following
expression
\begin{equation}
\mathbf{V}_{AB}^{\blacktriangleright}=\left(
\begin{matrix}
\lbrack\eta_{A}\mu+(1-\eta_{A})\gamma]\mathbf{I} & \sqrt{g\eta_{A}(\mu^{2}%
-1)}\mathbf{Z}\\
\sqrt{g\eta_{A}(\mu^{2}-1)}\mathbf{Z} & [g\mu+(g-1)\omega]\mathbf{I}%
\end{matrix}
\right)  , \label{VAB}%
\end{equation}
from which we can extract Alice's variance $V_{A}=\eta_{A}\mu+(1-\eta
_{A})\gamma$. Applying homodyne detection on mode $B$ we obtain the following
expression for Alice's variance conditioned to Bob outcomes
\begin{equation}
V_{A|\beta}=\frac{g\gamma\left(  1-\eta_{A}\right)  +\eta_{A}\omega\left(
g-1\right)  }{g}.
\end{equation}
From the expression of $V_{A}$ and $V_{A|\beta}$, and using the definition of
mutual information $I_{AB}=\frac{1}{2}\log_{2}V_{A}V_{A|\beta}^{-1}$, we
obtain the asymptotic Alice and Bob's mutual information, which is given by
\begin{equation}
I_{AB}\overset{\mu\rightarrow\infty}{=}\frac{1}{2}\log_{2}\frac{g\eta_{A}\mu
}{g\gamma(1-\eta_{A})+\eta_{A}\omega(g-1)}. \label{IAB-DR}%
\end{equation}

We then compute Eve's Holevo function, defined as $\chi_{AE}:=S_{T}%
-S_{C}^{\blacktriangleright}$, where $S_{T}$ is the von Neumann entropy of
$\rho_{E^{\prime}e}$, and $S_{C}$ is that of the conditional state
$\rho_{E^{\prime}e|A}$. We consider the block of CM $\mathbf{V}$ given by%
\begin{equation}
\mathbf{V}_{E^{\prime}eA}^{\blacktriangleright}=\left(
\begin{matrix}
\mathbf{V}_{E^{\prime}e}^{\blacktriangleright} & \mathbf{C}\\
\mathbf{C} & \mathbf{V}_{A}%
\end{matrix}
\right)  , \label{VEpeA}%
\end{equation}
where $\mathbf{V}_{A}=[\eta_{A}\mu+(1-\eta_{A})\gamma]\mathbf{I}$, and%
\begin{equation}
\mathbf{V}_{E^{\prime}e}^{\blacktriangleright}=\left(
\begin{matrix}
\lbrack(g-1)\mu+g\omega]\mathbf{I} & \sqrt{g(\omega^{2}-1)}\mathbf{Z}\\
\sqrt{g(\omega^{2}-1)}\mathbf{Z} & \omega\mathbf{I}%
\end{matrix}
\right)  ,~\mathbf{C}=\left(
\begin{matrix}
\sqrt{(g-1)\eta_{A}(\mu^{2}-1)}\mathbf{I}\\
0\mathbf{I}%
\end{matrix}
\right)  .
\end{equation}
We then compute the asymptotic symplectic spectrum of $\mathbf{V}_{E^{\prime
}e}^{\blacktriangleright}$, obtaining the following symplectic eigenvalues
\begin{equation}
\{\nu_{1},\nu_{2}\}\overset{\mu\rightarrow\infty}{\rightarrow}\{(g-1)\mu
,\omega\}.
\end{equation}
The total von Neumann entropy is $S_{T}=h(\nu_{1})+h(\nu_{2})$. Considering
that $h(x)=\log_{2}(ex/2)$ for $x\rightarrow\infty$, we can obtain the
following asymptotic formula
\begin{equation}
S_{T}\overset{\mu\rightarrow\infty}{=}\log_{2}\frac{e}{2}(g-1)\mu+h(\omega).
\label{ST-DR}%
\end{equation}
After Alice's homodyne detection of quadrature $q$ (or $p$) on mode $A$, we
also obtain Eve's conditional CM%
\begin{equation}
\mathbf{V}_{E^{\prime}e|A}^{\blacktriangleright}=\mathbf{V}_{E^{\prime}%
e}^{\blacktriangleright}-\mathbf{C}\left(  \Pi\mathbf{V}_{A}\Pi\right)
^{-1}\mathbf{C}^{T}, \label{V-Con-DR}%
\end{equation}
where $\Pi=\mathrm{diag}(1,0)$ for homodyne detection on $q$ and
$\Pi=\mathrm{diag}(0,1)$ for homodyne detection on $p$. From
Eq.~(\ref{V-Con-DR}) we can compute the symplectic spectrum of $\mathbf{V}%
_{E^{\prime}e|A}^{\blacktriangleright}$. After some algebra and working in the
limit of large modulation ($\mu\rightarrow\infty$), we obtain the analytical
expressions of the symplectic eigenvalues%
\begin{equation}
\bar{\nu}_{1}^{\blacktriangleright}\overset{\mu\rightarrow\infty}{=}%
\sqrt{\frac{(g-1)[g\eta_{A}\omega+\gamma(g-1)(1-\eta_{A})]}{\eta_{A}}\mu
,}~\bar{\nu}_{2}^{\blacktriangleright}\overset{\mu\rightarrow\infty}{=}%
\sqrt{\frac{\omega\lbrack g\eta_{A}+\gamma\omega(g-1)(1-\eta_{A})]}{g\eta
_{A}\omega+\gamma(g-1)(1-\eta_{A})}}.
\end{equation}

From this symplectic spectrum we can compute the conditional von Neumann
entropy $S_{C}=h(\bar{\nu}_{1})+h(\bar{\nu}_{2})$. For large $\mu$, it becomes%
\begin{equation}
S_{C}^{\blacktriangleright}\overset{\mu\rightarrow\infty}{=}\frac{1}{2}%
\log_{2}\frac{e^{2}}{4}\frac{(g-1)[g\eta_{A}\omega+\gamma(g-1)(1-\eta_{A}%
)]}{\eta_{A}}\mu+h\left(  \bar{\nu}_{2}^{\blacktriangleright}\right)
.\label{SC-DR}%
\end{equation}
Combining Eqs.~(\ref{ST-DR}) and~(\ref{SC-DR}) in the definition of the Holevo
function $\chi_{AE}:=S_{T}-S_{C}^{\blacktriangleright}$, we derive
\begin{equation}
\chi_{AE}\overset{\mu\rightarrow\infty}{=}\frac{1}{2}\log_{2}\frac
{(g-1)\eta_{A}\mu}{g\eta_{A}\omega+\gamma(g-1)(1-\eta_{A})}+h(\omega
)-h(\bar{\nu}_{2}^{\blacktriangleright}).\label{CHI-DR}%
\end{equation}
Finally, using Eqs. (\ref{IAB-DR}) and (\ref{CHI-DR}), we obtain the analytic
expression of the asymptotic key rate in DR, which is given in
Eq.~(\ref{Rate-DR}).~$\blacksquare$

The secret key rate of Eq. (\ref{Rate-DR}) can be optimized over Alice's free
parameters, which are the transmissivity $\eta_{A}\in\lbrack0,1]$ and the
variance $\gamma\geq1$. When $\eta_{A}=1$, which means we have no trusted
noise injected by Alice, it is easy to verify that $R^{\blacktriangleright
}(g,\omega,1,\gamma)=\log_{2}[g/(g-1)]-h(\omega)$, corresponding to the
previous lower bound $\Omega$ in Eq.~(\ref{LBonly}). It is therefore clear
that the optimized achievable rate $R^{\blacktriangleright}$ in
Eq.~(\ref{Bound-DR}) is $\geq\Omega$ for any value of the gain. In the
numerical comparison below (subsection~\ref{SecCOMP}) we explicitly show that
there is a strict separation, so that we have $R^{\blacktriangleright}>\Omega$
in a wide range.

\subsection{Achievable rate in reverse reconciliation\label{SECRR}}

We now show the following.

\begin{theorem}
Consider a thermal amplifier channel with gain $g$ and thermal noise $\omega$.
Its secret key rate is lower-bounded by the achievable RR\ rate
\begin{equation}
R^{\blacktriangleleft}(g,\omega)=\underset{\eta_{B},\gamma}{\max
}~R^{\blacktriangleleft}\left(  g,\omega,\eta_{B},\gamma\right)
,\label{Bound-RR}%
\end{equation}
where
\begin{equation}
R^{\blacktriangleleft}(g,\omega,\eta_{B},\gamma):=\frac{1}{2}\log_{2}%
\frac{\eta_{B}\omega+\gamma(g-1)(1-\eta_{B})}{(g-1)[\gamma(1-\eta_{B}%
)+\eta_{B}\omega(g-1)]}+h\left(  \sqrt{\frac{\omega\lbrack\eta_{B}%
-\gamma\omega(g-1)(1-\eta_{B})]}{\eta_{B}\omega+\gamma(g-1)(1-\eta_{B})}%
}\right)  -h(\omega),\label{Rate-RR}%
\end{equation}
and the maximization is over the transmissivity $\eta_{B}$ of Bob's beam
splitter, and the thermal variance $\gamma\geq1$.
\end{theorem}

\begin{figure}[th]
\vspace{-2cm} \centering\includegraphics[width=13cm]{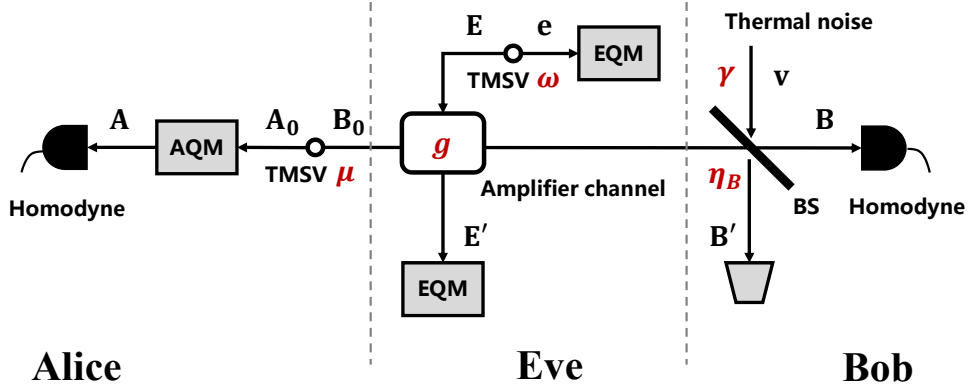}
\vspace{-1.55cm} \caption{Protocol with trusted thermal noise in
RR. Alice starts from a TMSV where mode $B_{0}$ is sent to Bob
through a thermal amplifier channel with gain $g$ (and thermal
noise $\omega$), while mode $A_{0}$ is stored in Alice's quantum
memory (AQM), waiting for the correct sequence of homodyne
detections, which is announced by Bob at the end of the protocol.
The attack is assumed to be collective, with Eve using a TMSV
state, whose output modes, $e$ and $E^{\prime}$, are stored in her
quantum memory (EQM). At the channel output, the signals are
processed within Bob's private space, by a beam splitter (BS) with
transmissivity $\eta_{B}$ and a thermal mode $v$ with variance
$\gamma$. The signal modes are then homodynes either in $q$ or
$p$. At the end, Bob publicly declares to Alice his sequence of
homodynes. At this point, Alice performs the correct sequence of
homodyne detections on the modes $A$ she stored in her quantum
memory.}%
\label{Structure-RR}%
\end{figure}

\textbf{Proof.} The proof is similar to the DR discussed in previous section.
Consider the Gaussian protocol in Fig.~\ref{Structure-RR}. Alice starts from
the same TMSV state $\Phi^{\mu}$, of modes $A_{0}$ and $B_{0}$, given in
Eq.~(\ref{VA0B0}). Now, it is Alice's mode $A$ that is stored in a Alice's
quantum memory (AQM) for later measurements, while mode $B_{0}$ travels to Bob
through the amplifier channel. Bob implements a noisy detection, mixing the
input mode with a thermal mode $v$ with variance $\gamma$ via a beam splitter
whose transmissivity is $\eta_{B}$. Then Bob measures the $q$ or the $p$
quadrature (communicating its choices at the end of the quantum communication
after $n\gg1$ rounds).

The initial global state of Alice, Bob and Eve has CM $\mathbf{V}%
_{0}^{\blacktriangleleft}=\mathbf{V}_{AB_{0}}\mathbf{\oplus V}_{eE}%
\mathbf{\oplus V}_{v}=\mathbf{V}_{A_{0}B_{0}eEv}^{\blacktriangleleft}$, and we
again rearrange the modes so that $\mathbf{V}_{0}^{\blacktriangleleft
}=\mathbf{V}_{A_{0}vB_{0}Ee}^{\blacktriangleleft}$. This state is processed by
the amplifier ($g$) and then the beam splitter ($\eta_{B}$). First we obtain
$\mathbf{\tilde{V}}^{\blacktriangleleft}=\mathbf{S}_{g}\mathbf{V}%
_{0}^{\blacktriangleleft}\mathbf{S}_{g}^{T}$, where $\mathbf{S}_{g}$ has been
defined above, and then we compute $\mathbf{V}^{\blacktriangleleft}%
=\mathbf{S}_{\eta_{B}}\mathbf{\tilde{V}}^{\blacktriangleleft}\mathbf{S}%
_{\eta_{B}}^{T}$, where $\mathbf{S}_{\eta_{B}}:=\mathbf{I\oplus T}_{BS}%
(\eta_{B})\oplus\mathbf{I\oplus I}$, with $\mathbf{T}_{BS}(\cdot)$ and
$\mathbf{T}_{AMP}(\cdot)$ given in Eq.~(\ref{T-BS}) and (\ref{T-AMP}).
Discarding Bob's mode $B^{\prime}$, we compute the output state $\rho
_{ABE^{\prime}e}^{\blacktriangleleft}=\mathrm{Tr}_{B^{\prime}}(\rho
_{AB^{\prime}BE^{\prime}e})$ with CM $\mathbf{V}_{ABE^{\prime}e}%
^{\blacktriangleleft}$. From this CM we may compute the secret key rate in RR,
i.e.,%
\begin{equation}
R^{\blacktriangleleft}=I_{AB}-\chi_{BE},\label{rate-RR-DEF}%
\end{equation}
where $\chi_{BE}$ is Eve's Holevo information on Bob's outcomes. The
analytical form of $R^{\blacktriangleleft}$ is obtained under the assumption
of large modulation $\mu\rightarrow\infty$.

The CM describing Alice's and Bob's output modes $A$ and $B$ is the following
\begin{equation}
\mathbf{V}_{AB}^{\blacktriangleleft}=\left(
\begin{matrix}
\mu\mathbf{I} & \sqrt{g\eta_{B}(\mu^{2}-1)}\mathbf{Z}\\
\sqrt{g\eta_{B}(\mu^{2}-1)}\mathbf{Z} & [\eta_{B}(g\mu+(g-1)\omega
)+(1-\eta_{B})\gamma]\mathbf{I}%
\end{matrix}
\right)  .
\end{equation}
We can therefore compute Alice and Bob's mutual information
\begin{equation}
I_{AB}^{\blacktriangleleft}\overset{\mu\rightarrow\infty}{=}\frac{1}{2}%
\log_{2}\frac{g\eta_{B}\mu}{\gamma(1-\eta_{B})+(g-1)\eta_{B}\omega
}.\label{IAB-RR}%
\end{equation}
Eve's Holevo information can be written as $\chi_{BE}=S_{T}%
^{\blacktriangleleft}-S_{C}^{\blacktriangleleft}$, where $S_{T}%
^{\blacktriangleleft}$ is the von Neumann entropy for Eve's total state
$\rho_{E^{\prime}e}$, while $S_{C}^{\blacktriangleleft}$ is obtained from the
conditional quantum state $\rho_{E^{\prime}e|B}$. For its computation,
consider the following CMs
\begin{equation}
\mathbf{V}_{BE^{\prime}e}^{\blacktriangleleft}=\left(
\begin{matrix}
\mathbf{V}_{E^{\prime}e}^{\mathbf{\blacktriangleleft}} & \mathbf{\bar{C}}\\
\mathbf{\bar{C}} & \mathbf{V}_{B}%
\end{matrix}
\right)  ,~\mathbf{V}_{E^{\prime}e}^{\blacktriangleleft}=\left(
\begin{matrix}
\lbrack(g-1)\mu+g\omega]\mathbf{I} & \sqrt{g(\omega^{2}-1)}\mathbf{Z}\\
\sqrt{g(\omega^{2}-1)}\mathbf{Z} & \omega\mathbf{I}%
\end{matrix}
\right)  ,~\mathbf{\bar{C}}=\left(
\begin{matrix}
\sqrt{g(g-1)\eta_{B}}(\mu+\omega)\mathbf{Z}\\
\sqrt{(g-1)\eta_{B}(\omega^{2}-1)}\mathbf{I}%
\end{matrix}
\right)  ,
\end{equation}
where $\mathbf{V}_{B}=[\eta_{B}(g\mu+(g-1)\omega)+(1-\eta_{B})\gamma
]\mathbf{I}$\textbf{. }Clearly, we need to compute only Eve's conditional
symplectic spectrum, obtained from Eve and Bob's CM, $\mathbf{V}_{BE^{\prime
}e}^{\blacktriangleleft}$ by applying homodyne detection on Bob mode $B$. This
provides the conditional CM
\begin{equation}
\mathbf{V}_{E^{\prime}e|B}^{\mathbf{\blacktriangleleft}}=\mathbf{V}%
_{E^{\prime}e}^{\mathbf{\blacktriangleleft}}-\mathbf{\bar{C}}\left(
\Pi\mathbf{V}_{B}\Pi\right)  ^{-1}\mathbf{\bar{C}}^{T},\label{V-Con-RR}%
\end{equation}
whose symplectic eigenvalues have the following asymptotic expressions%
\begin{equation}
\bar{\nu}_{1}^{\mathbf{\blacktriangleleft}}\overset{\mu\rightarrow\infty}%
{=}\sqrt{\frac{(g-1)[\eta_{B}\omega+\gamma(g-1)(1-\eta_{B})]}{g\eta_{B}}\mu
},~\bar{\nu}_{2}^{\mathbf{\blacktriangleleft}}\overset{\mu\rightarrow\infty
}{=}\sqrt{\frac{\omega\lbrack\eta_{B}+\gamma\omega(g-1)(1-\eta_{B})]}{\eta
_{B}\omega+\gamma(g-1)(1-\eta_{B})}}.
\end{equation}

Therefore, the asymptotic Eve's Holevo information is given by
\begin{equation}
\chi_{BE}\overset{\mu\rightarrow\infty}{=}\frac{1}{2}\log_{2}\frac{(g-1)\mu
}{\eta_{B}\omega+\gamma(g-1)(1-\eta_{B})}+h(\omega)-h(\bar{\nu}_{2}%
).\label{CHI-RR}%
\end{equation}
Combining Eqs.~(\ref{IAB-RR}) and (\ref{CHI-RR}), in Eq.~(\ref{rate-RR-DEF}),
we find the formula of the asymptotic key rate in RR in the asymptotic limit
of large Gaussian modulation, which coincides with that given in
Eq.~(\ref{Rate-RR}). The secret key rate of Eq.~(\ref{Rate-RR}) is then
optimized over Bob's free parameters $\eta_{B}\in\left[  0,1\right]  $ and
$\gamma\geq1$.~$\blacksquare$

\subsection{Comparison\label{SecCOMP}}

\begin{figure}[th]
\centering \subfigure{
\includegraphics[width=8cm]{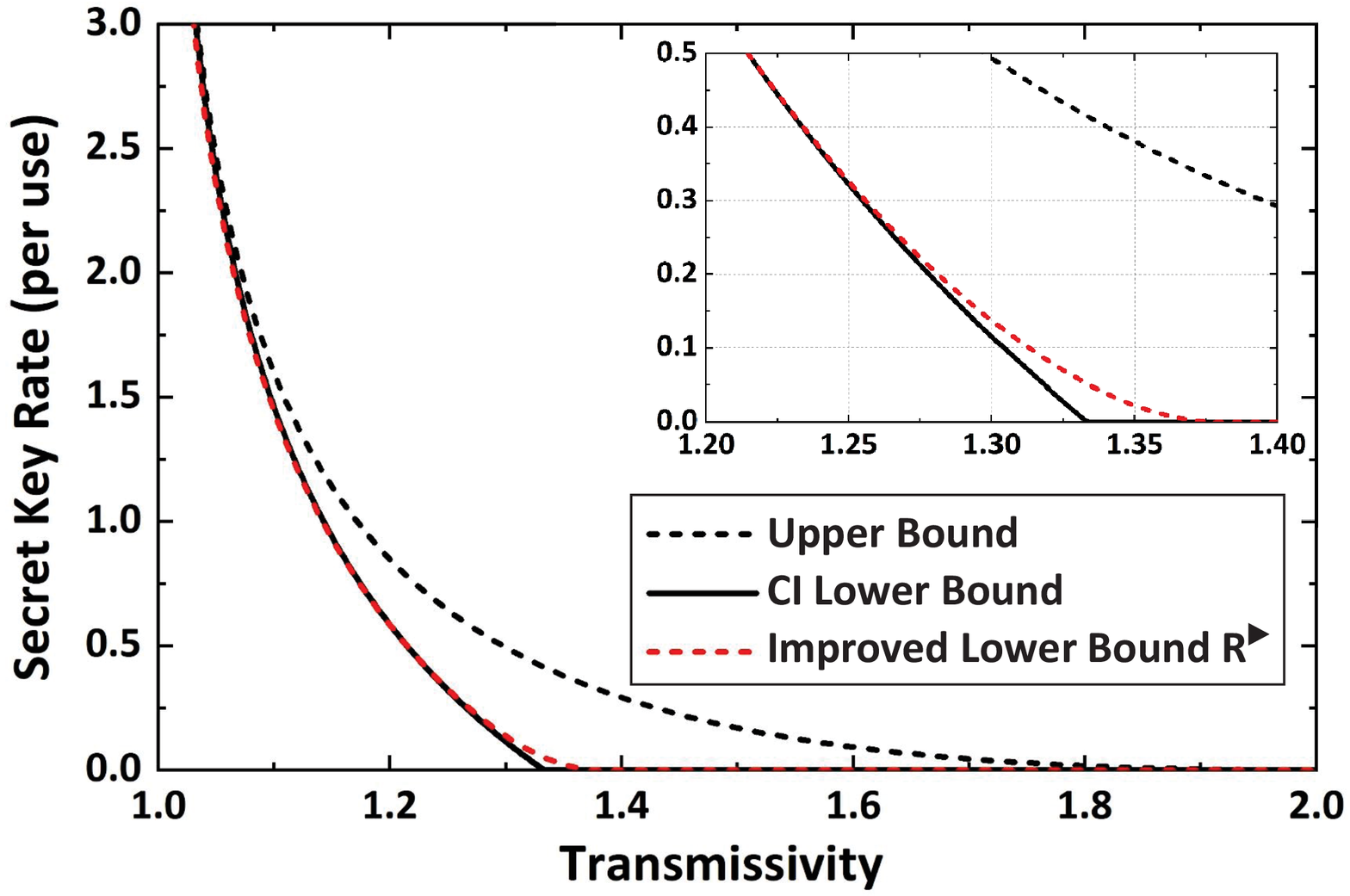}
\includegraphics[width=8cm]{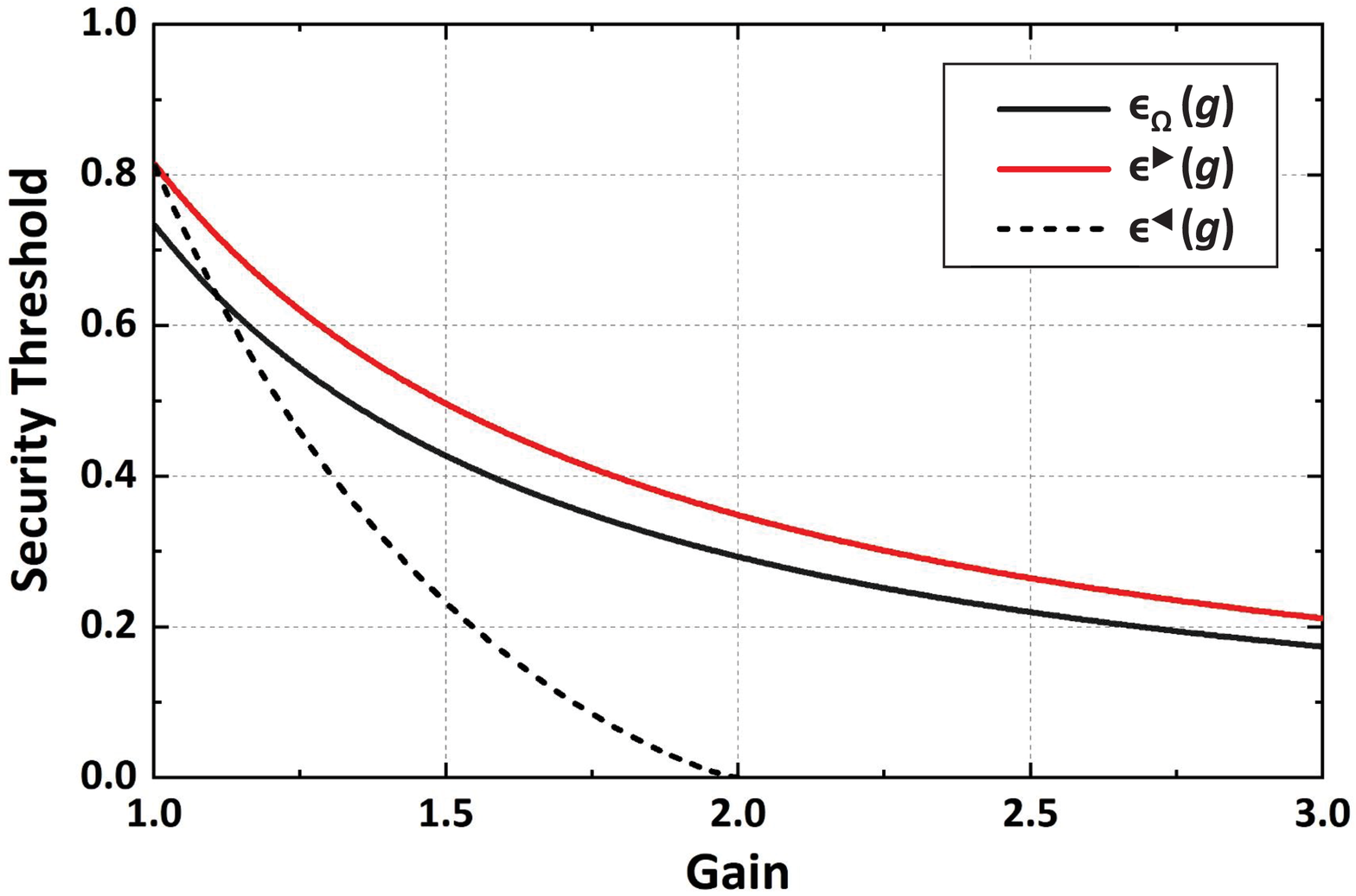}
} \caption{Comparison between the previous lower bound and the new improved
results. In the left panel, we consider a thermal amplifier channel with
$\bar{n}=1$ photons and arbitrary gain $g$. We then compare the new lower
bound $R^{\blacktriangleright}$ of Eq.~(\ref{Bound-DR}) (red-dashed line) with
the coherent information of the channel $\Omega$ of Eq~(\ref{LBonly})
(solid-black line). We also show the upper bound $\Phi$ of Eq.~(\ref{UBonly})
(black-dashed line). In the right panel, we compare the security thresholds
$\epsilon_{\Omega}(g)$ (black-solid line), $\epsilon^{\blacktriangleright}(g)$
(red-solid line),\ and $\epsilon^{\blacktriangleleft}(g)$ (black-dashed
line).}%
\label{comparison}%
\end{figure}

The performances of the new lower bounds are summarized in
Fig.~\ref{comparison}. The left panel compares the improved lower bound in the
DR $R^{\blacktriangleright}$ of Eq.~(\ref{Bound-DR}) (red-dashed line) with
respect to the previous lower bound $\Omega$ of Eq.~(\ref{LBonly}) given by
the coherent information of the channel (black-solid line). We also show the
upper bound $\Phi$ of Eq.~(\ref{UBonly}) denoted by the black-dashed line.
Then, we compare the security thresholds in the right panel of
Fig.~\ref{comparison}. Let us define the excess noise of the thermal amplifier
channel as $\epsilon=(g-1)(\omega-1)/g$. Then, we may write the rates as
$R=R(g,\epsilon)$. Setting $R=0$, we therefore find the maximally-tolerable
excess noise as a function of the gain, i.e., $\epsilon=\epsilon(g)$. Starting
from $\Omega$ and the two optimized rates $R^{\blacktriangleright}$ and
$R^{\blacktriangleleft}$, we therefore compute the corresponding security
thresholds $\epsilon_{\Omega}(g)$, $\epsilon^{\blacktriangleright}(g)$ and
$\epsilon^{\blacktriangleleft}(g)$ which are plotted in the right panel of
Fig.~\ref{comparison}. As we can see, $\epsilon^{\blacktriangleright
}(g)>\epsilon_{\Omega}(g)$ for any $g$, while $\epsilon^{\blacktriangleleft
}(g)$ outperforms $\epsilon_{\Omega}(g)$ only for smal gains.

\section{Conclusions}

In this work, we have studied a QKD protocol whose rate is able to improve the
lower bound to the secret-key capacity of the thermal amplifier channel. In
DR\ this improvement occurs for any value of the gain $g$ and the thermal
noise $\bar{n}$ of the channel. Our protocol is based on randomly-switched
squeezed states and homodyne detections, which are perfectly reconciliated by
resorting to a quantum memory. Most importantly, we employ a beam-splitter and
trusted thermal noise just before the homodyne detector. The large-modulation
($\mu\rightarrow\infty$) and asymptotic ($n\rightarrow\infty$) secret key rate
is then optimized over the free parameters of the over-all noisy detection.
Even though the gap between the new lower bounds and the upper bound is still
quite large, our work confirms the fact that the coherent information of the
thermal amplifer channel is well separated from its secret key capacity. This
also seems to suggest that the distribution of secret keys over this quantum
channel might occur at higher rates than the distribution of entanglement or
the transmission of quantum information.

\acknowledgments

CO and SP acknowledge support from the EPSRC via the `Quantum Communications
HUB' (EP/M013472/1). GW and HG acknowledge support from the National Natural
Science Foundation of China (Grant No. 61531003).

\end{document}